\documentstyle[11pt,IAU207_pasp,twoside,psfig]{article}
\def\edcomment#1{\iffalse\marginpar{\raggedright\sl#1\/}\else\relax\fi}
\reversemarginpar

\begin{document}
\title{ New Insights from a study of the Globular
       Cluster Systems of 60 Early-Type Galaxies}
 \author{Arunav Kundu}
\affil{Department of Astronomy, Yale University, 260 Whitney Ave, New Haven, CT 06511, U.S.A.}
\author{Brad Whitmore}
\affil{Space Telescope Science Institute, 3700 San Martin Drive, Baltimore, MD 21218, U.S.A.}

\begin{abstract}
We  present the results of our detailed WFPC2-based photometric
study of the globular cluster systems (GCS) of over 60 elliptical and S0 galaxies. Approximately 50\% of the
GCSs of ellipticals, and at least 15-20\% of S0s reveal bimodality in the color distribution. We trace the variation in GCS properties with Hubble type and 
discuss the implications  on the various models of galaxy 
(and cluster system) formation. We also provide evidence that the globular
cluster luminosity function is an excellent distance indicator with
an accuracy comparable to the surface brightness fluctuation method.

\end{abstract}

\section{Introduction}

	Globular clusters (GC) are among the most pristine objects in galaxies and hence  provide
 invaluable insights into the chemical and kinematic conditions prevalent
 during the early epochs of galaxy formation. Just how early is presently 
a matter of some debate. Initial studies of the GCSs
of the Milky Way and other nearby galaxies led to the conclusion
 that all globular
clusters are old, metal-poor systems that were formed during
the initial collapse phase of a protogalactic gas cloud. The subsequent 
discovery of metal-rich GCSs and bimodal color (metallicity) distributions in some 
galaxies has led to much discussion about alternate mechanisms of globular cluster 
formation, including at this very symposium. Schweizer (1987) and Ashman $\&$
 Zepf (1992) proposed that some GCs may be formed during the interaction or merger
of galaxies. While Zepf $\&$ Ashman (1993) suggest that the bimodality in the color
distribution is a natural consequence of the merger scenario, Forbes, Brodie \&
Grillmair (1997) favor multiple phases of cluster formation during the 
collapse of a gas cloud into a galaxy.

 	 Only recently has it become possible to test these and other  theories of 
globular cluster formation and evolution, due to the explosion in the number 
of observations of GCSs of external galaxies.
 The advent of the HST with its superior angular resolution has given a major boost to the field because of the ease with which cluster candidates can be identified.  While most HST-based studies concentrate on a single galaxy, or just a handful, there are  few integrated analyses that study a  large number of galaxies simultaneously (Forbes et al. 1996; Gebhardt \& Kissler-Patig  1999; Brodie \& Larsen elsewhere in this volume) which helps eliminate the possible systematic
 differences that make comparison of the results of different authors a
somewhat tricky task. We report here the results from our recent analysis of the GCS of over 60 S0 and elliptical galaxies using WFPC2 V and I images (Kundu \& Whitmore 2001a; 2001b). This is the largest and most comprehensive analysis of GCSs to date.

\section {Color Distributions and Implications on Formation Models}

\begin{figure}[!ht]
\centerline{\psfig{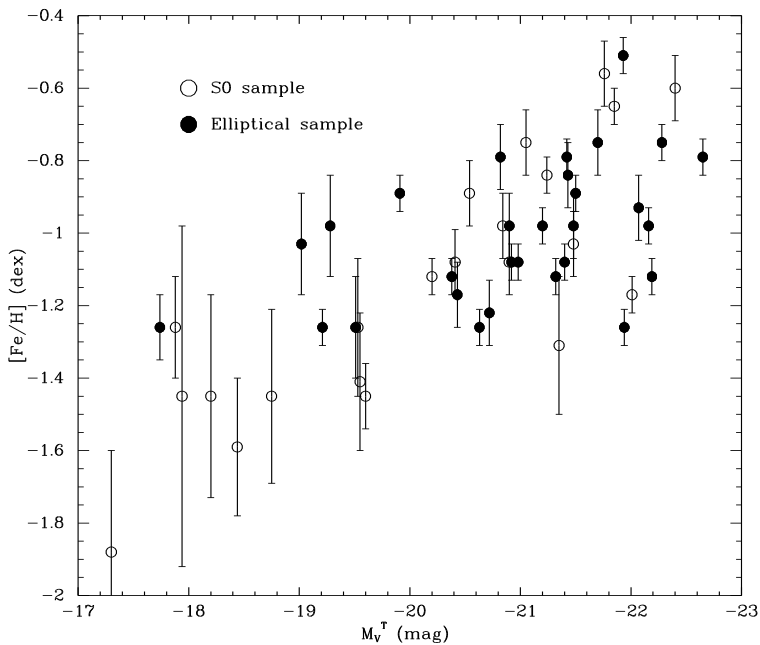}}
\phantom{a}
\caption{The average metallicity of the cluster systems  vs the absolute magnitude of the host galaxy. Systems with too few clusters for reliable mean color estimates have not been plotted.}
\end{figure}

Our analysis of the WFPC2 images of the inner regions of the 29 elliptical galaxies in our sample revealed a measurable GCS in each of them (Kundu \& Whitmore  2001a). However, only 30 of the 35 S0s (Kundu \& Whitmore 1998; 2001b) showed evidence of a GCS. This may be partly due to the fact that S0s have intrinsically fewer clusters than ellipticals (see
 \S4), and partially because for 34 of the 35 S0 candidates we analyzed were short exposure 'snapshot' images. The  majority of  GC candidates 
  lie in a narrow range of color  between 0.5$<$V-I$<$1.5 with 
a mean color near V-I $\approx$ 1.0 mag, which is fairly typical for old globular cluster systems. Hence, the color of the cluster is predominantly a function of the metallicity, and throughout this analysis we use the linear color-metallicity relationship for Milky Way GCs (Kundu \& Whitmore 1998) to quote metallicities (see Kissler-Patig in this volume for an alternate view).

 The mean color of the GCSs of our entire elliptical sample is 1.04$\pm$0.04 (0.01) which is  indistinguishable from the value of
1.00$\pm$0.07 (0.01) for the S0s. \footnote {We  use the following convention in quoting uncertainties: The number following the $\pm$ sign is the standard deviation, while a number in parentheses refers to the uncertainty in the mean.}
Therefore, inasmuch as the broad-band colors trace metallicity, we can conclude that the metallicity of the GCS of early type galaxies is not a function of Hubble type.  What then drives this metallicity/color spread?  It has been suggested that that the mean metallicities  of elliptical galaxy GCSs are 
correlated to the mass of the host.  To check the veracity of this claim we plot the mean color of the elliptical and S0 GCSs vs the absolute magnitude (mass) of the host galaxies  in Fig 1.  We find that [Fe/H] generally increases with galaxy luminosity, and that there is no offset between the the S0 and elliptical samples. While there are no low luminosity S0s or ellipticals with high mean metallicities, there appear to be a number of luminous elliptical galaxies that have metallicities significantly 
lower than that expected for a linear [Fe/H]-M$_V^T$ relationship. This  difference in the mean metallicities of the cluster systems of hosts with similar luminosities may
  either be due to an offset in the entire metallicity distribution of the two 
cluster systems, or differing fractions of two or more sub-populations having 
different metallicities. To this end it is critical to study the detailed color 
distributions of individual GCSs. 

\subsection {Bimodality in Color Distributions}

	Ashman \& Zepf (1992) predicted bimodality in GCS color distributions based on their merger model, and subsequently observed it (Zepf \& Ashman 1993). 
While bimodality has been confirmed in a large number of elliptical galaxies since then, the frequency of bimodality and it's correlation with various host properties, which is central to the various competing theories of GCS formation, has not been very well studied. 

\begin{figure}[!ht]
\centerline{\psfig{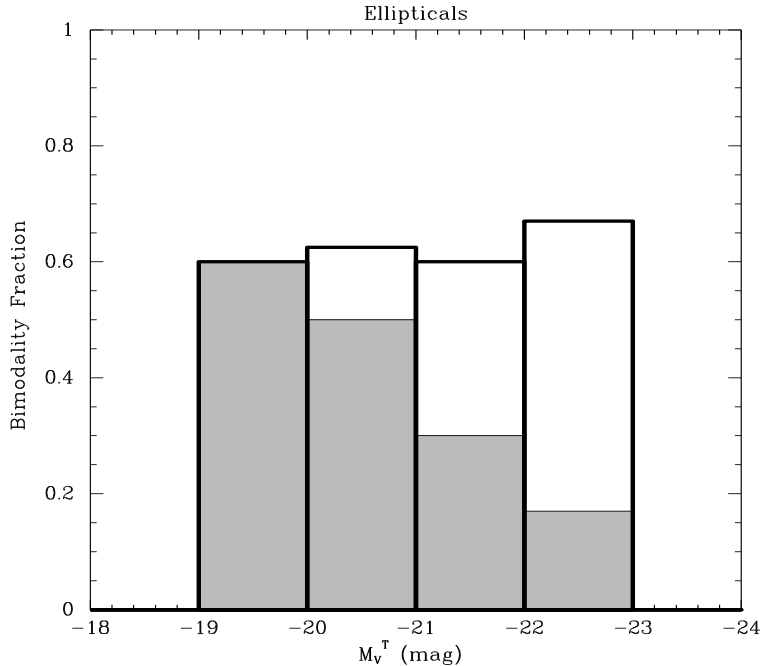}}
\caption{ Bimodality fraction in ellipticals as a function of host galaxy absolute magnitude. Unshaded areas represent galaxies with confirmed bimodality while the shaded histograms represent 'very likely' bimodal systems. The bimodalilty fraction is similar at all host galaxy absolute magnitudes. These fractions are likely to be lower limits. }
\end{figure}

	We searched for bimodality in our galaxy sample, using the KMM mixture modelling algorithm (Ashman, Bird \& Zepf 1994), fitting Gaussians to binned data, and visual inspection. The galaxies were sorted into three broad categories, confirmed bimodal, very likely bimodal, and those with no statistically measurable evidence of bimodality. Our analysis  revealed that 18 of the 29 ellipticals  can be better described by two Gaussian sub-populations than a single Gaussian, of which 9 are confirmed as bimodal without a doubt. The other 11 showed no reasonable partitions for two or more groups. As showed in Fig 2, roughly 60\% of ellipticals at all host masses are likely to be bimodal. This number is probably an underestimate as incidences of multiple populations may be difficult to discern due to photometric uncertainties or age-metallicity degeneracies. None of the GCSs in our sample show meaningful partitions for three or more groups. 

 Due to the short exposure times of the 'snapshot' S0 images and the smaller number of clusters per galaxy (\S4) we were unable to conclusively test for multiple populations of GCs in most of our S0s. Although only 2 of the S0s  revealed definitive evidence for bimodality, statistical tests  suggest that at least 15-20\% of these galaxies are bimodal at the present level of photometric accuracy.

\subsection {Implications on Formation Models}

	There are three competing models/scenarios that attempt
to explain  bimodality in GCS colors. While the merger model (Schweizer 1987; Ashman $\&$ Zepf 1992) and the  multiple collapse model (Forbes, Brodie, \& Grillmair 1997) both suggest that this is the consequence of globular clusters forming during two distinct epochs in the metal enrichment histories of these galaxies, the C\^{o}t\'{e} et al.  (1998) model proposes that all cluster systems formed at 
roughly the same time and that the bimodal distribution is a result of the merger (without creation of new clusters) or cannibalism of a metal-poor GCS by a larger galaxy with a pre-existing metal-rich GCS.

\begin{figure}[!ht]
\centerline{\psfig{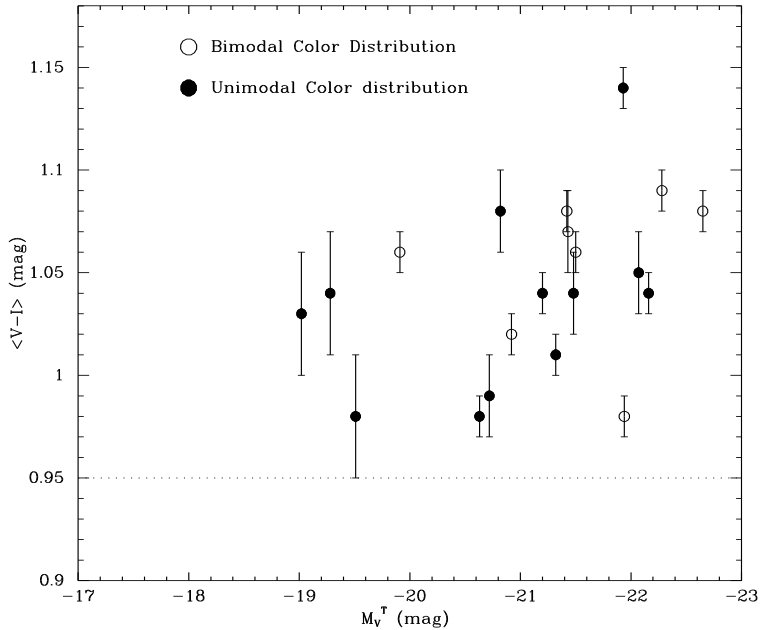}}
\caption{The mean color of the GCSs in ellipticals with confirmed bimodality or confirmed lack of bimodality as a function of host galaxy absolute magnitude. No significant  differences are observed between the two populations. }
\end{figure}

	In Fig 3 we plot the mean GCS colors of the galaxies with confirmed bimodal distributions and the ones with confirmed lack of bimodality as a function of host luminosity. Both distributions appear indistinguishable with
no evidence that the unimodal galaxies follow any kind of well defined  color(metallicity)-luminosity relation. To recall, the  C\^{o}t\'{e} et al.  (1998)  model assumes that the input unimodal galaxies that are cannibalized to form giant ellipticals have a second order luminosity-cluster metallicity relationship.   A set of merging/coalescing cluster systems with widely varying metallicity distributions, as seen in Fig 3, is  unlikely to produce a bimodal distribution in the absence of such a relation, which suggests that the  C\^{o}t\'{e} et al.  (1998) model is unviable. Moreover,  within the range of host luminosities of our sample we find no unimodal cluster system with a mean GCS color of V-I$\approx$1.2 (corresponding to the red peak of giant galaxies) which could serve as the progenitor host for the C\^{o}t\'{e} et al. cannibalism model.  We suspect that the reason for the shallower input 
color-luminosity relationship  used by C\^{o}t\'{e} et al.  (1998)  is that
they calculated it on the basis of ground-based data. It has been shown that in 
galaxies with bimodal color distributions the blue clusters are more spatially
extended than the red ones (e.g. Geisler, Lee \& Kim 1996), so it is very likely that a relationship derived on the basis of ground-based data, that mostly samples the outer regions of a cluster system, returns a shallower color-luminosity relationship. The low luminosity elliptical galaxies present
a further problem for the C\^{o}t\'{e} et al model. In order to create bimodal low mass systems (See Fig 2) through the 'cannibalism' model one 
would require a population of low mass, unimodal metal-rich systems, which could then accrete other low mass metal-poor systems. There are two problems with this hypothesis: We observe no low mass metal-rich systems in our sample. And such
a requirement would violate the smooth second order metallicity(color)-host luminosity relationship required for the formation of bimodal systems in giant galaxies through the  C\^{o}t\'{e} et al model. 
Thus the observed bimodality
 in low luminosity galaxy could not have been caused by the mergers of two or
more systems of different metallicities via the 'cannibalism' model, and some other mechanism is at work. If
such an independent method for creating bimodality exists in low luminosity galaxies, one could turn this argument around and claim that the same process
could occur in giant galaxies, and that the accreting low mass galaxies bring bimodal populations with them.  While it is very likely that giant galaxies cannibalize dwarfs and their  predominantly metal-poor cluster systems to some extent, it is unlikely to be the primary mechanism of creating bimodality in a majority of galaxies. Thus, our observations do not seem to support the C\^{o}t\'{e} et al.  (1998) model and suggests that the bimodality in globular cluster colors is evidence for the formation of clusters in two different 
epochs of the metal-enrichment process. By extension this implies that for old cluster systems the red clusters are younger than the blue ones.

	There are no obvious ways to distinguish between the multiple collapse
and merger models based on the colors of clusters. They both predict bimodal color distributions with the red clusters situated closer to the center of the
 galaxy than the blue ones (which is observed in our sample). While the merger model provides a mechanism for the second episode of cluster formation, one of the drawbacks of the multiple collapse model is that at present there is no known viable trigger. It is  possible that the second burst of cluster formation in the collapse model is 
triggered by a merger.  The massive gas budget required to form the many thousands of metal-rich GCs in giant ellipticals, coupled with the roughly similar inferred ages of the red and blue clusters observed in galaxies like M87,
 suggests that for the merger model to work the event must have taken place very early in the history of the progenitors when they were still gaseous. The difference between the merger of two largely gaseous bodies and the collapse of
one large gaseous entity (with possible  fragments within this body) may largely
be one of semantics.

	In NGC 3115, the only S0 with deep observations and sufficient number of clusters for independent analysis of the red and blue GCS, we find that the red
clusters are associated with the thick disk while the blue ones are members of
a spheroidal population. This is consistent with the minor merger scenario of Schweizer (1990). If the metal-poor GCs were formed during the initial collapse
phase of the galaxy, the merger of a smaller galaxy (mass ratio $\sim$0.3) could cause a starburst in the disk that heats the gas and induces star formation in giant molecular clouds, most of which are likely to be associated with the disk of the larger of the two merging galaxies.

	Based on our analysis of early type galaxies we conclude that a majority of the GCSs 
of ellipticals are formed in two episodes. The second episode is 
likely triggered by a merger; possibly a major merger in the case of cD galaxies
like M87 and a minor merger in the case of S0 galaxies like NGC 3115 that
preserves the disk component.

\section{The Globular Cluster Luminosity Function}

\begin{figure}
\centerline{\psfig{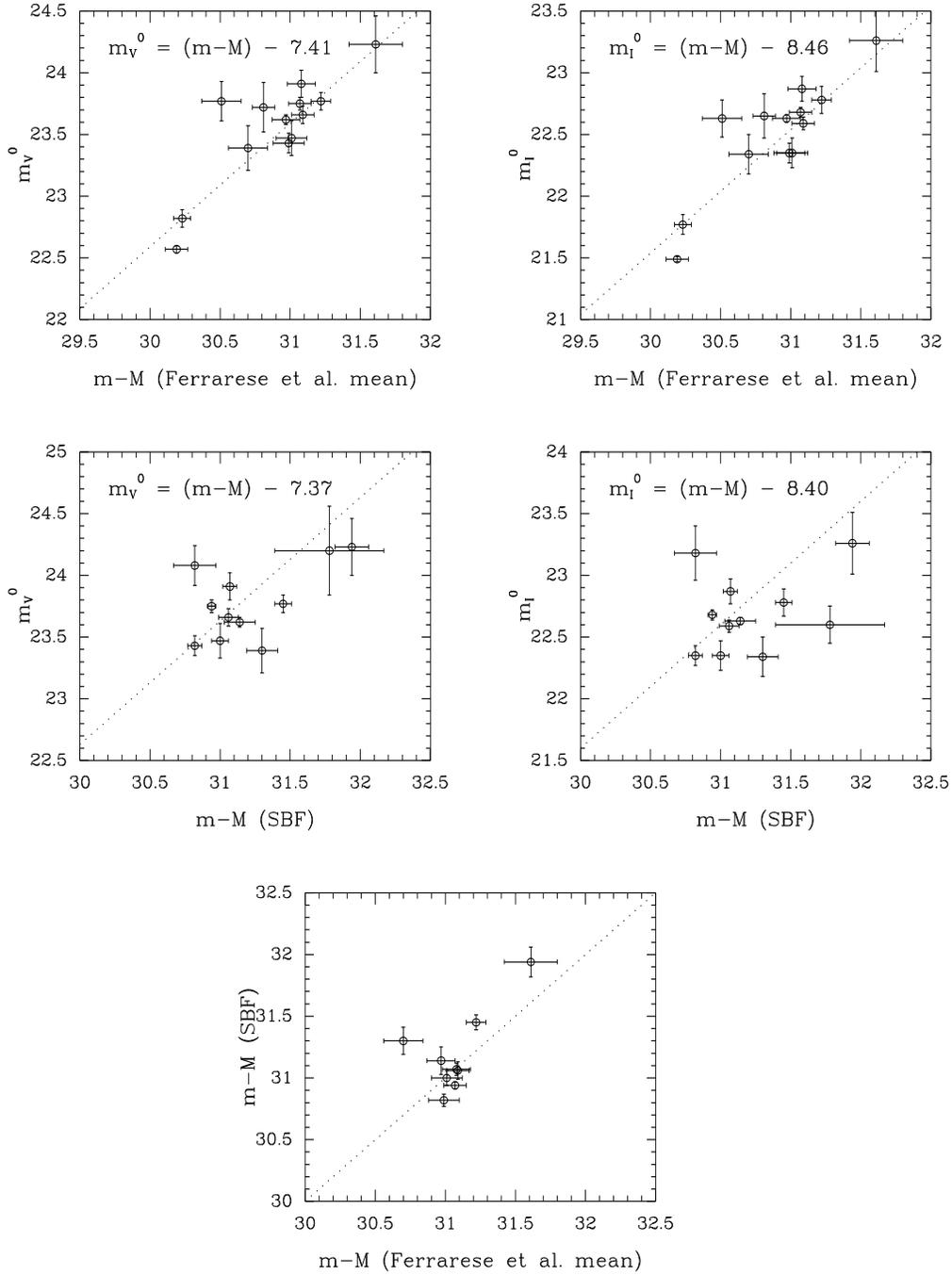}}
\caption{Top: GCLF turnovers in the V and I-bands  compared with the distance
moduli from Ferrarese et al. (2000). The dotted lines trace the constant offset between the turnover and distance moduli. Middle: Corresponding comparisons with the Neilsen et al. (1999) SBF distance measurements. Bottom: SBF measurements (Neilsen et al.)  vs the Ferrarese et al. mean distance moduli. The GCLF turnover appears to be an excellent distance indicator with an
accuracy comparable with the SBF method. }
\end{figure}

	The turnover luminosity of the globular cluster luminosity function (GCLF) has been found to be remarkably constant over a wide range of galaxies and environments. So constant in fact that it has been used as a secondary distance indicator (Harris 1991, Jacoby 1992). The theoretical basis for this rather remarkable
result - which implies that for any reasonable range of M/L ratio the underlying
mass distribution of globular clusters is the same in all galaxies - is not well
understood.

	After detailed corrections for completeness effects, contamination etc. we fitted Gaussians curves to the GCLFs of our sample. While we found weak evidence that the width of the GCLF i.e. the dispersion increases with host mass, the variation in the dispersions is smaller than that reported in the literature. Overall a Gaussian with $\sigma$=1.3 mag describes both V and I band GCLFs quite well. The large scatter in previous observations can probably be attributed to the larger uncertainties induced by incompletenes and contamination in ground-based analyses. 

\begin{figure}
\centerline{\psfig{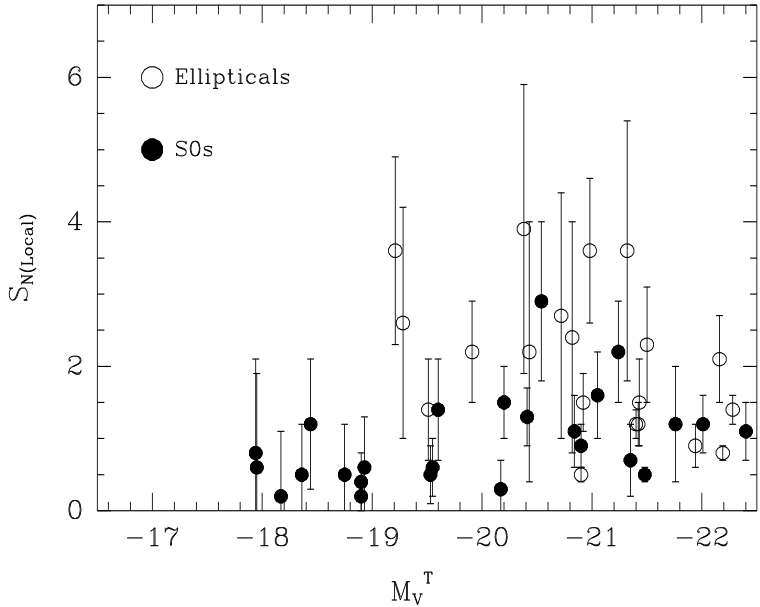}}
\caption{The variation of the local specific frequency within our field of view with
host galaxy magnitude.  The uncertainty in the local specific frequency of S0s is underestimated by a factor of 2 as compared to the ellipticals due to the different reduction methods applied to the short exposure S0 images. }
\end{figure}

In Fig 4 we compare the GCLF turnover luminosities of the GCLFs of the 11 systems with the deepest data in the V and I bands with the weighted distance moduli of three other distance indicators from Ferrarese et al. (2000) and
the surface brightness fluctuation distances from Neilsen (1999).  We also compare the Neilsen distances to
the weighted means. The GCLF turnover, in both the V and I-band tracks the distance to these galaxies very well throughout the entire distance range and appears to be an excellent distance indicator. Comparing our turnover magnitudes with the Ferrarese et al. distances we calculate a turnover luminosity of M$_V^0$ = -7.41(0.03) in V, and 
M$_I^0$=-8.46 (0.03) in the I-band.
 We also find evidence that the difference in the turnover luminosities in V and I increases with metallicity, as predicted by Ashman, Conti \& Zepf  (1995). In
sum we conclude that the GCLF is an excellent distance indicator with an 
accuracy that is as good as the surface brightness fluctuation method.

\section {Specific Frequency}

The specific frequency (S$_N$) is a measure of the luminosity normalized globular cluster density in a host galaxy, and is typically measured for the entire cluster system. Due to the limited field of view of the WFPC2 we could only measure the local value of S$_N$ in the inner regions of our galaxies. Fig 5 plots S$_{N(local)}$ for the S0s and ellipticals with reliable estimates as a function of the host galaxy luminosity. The reduction methods used to analyze the 'snapshot' S0 images were different from those used for the deeper ellipticals; hence the uncertainties are underestimated for the former by a factor of $\sim$2. However, it is clear that at all host masses ellipticals are more cluster rich than S0s. The mean  S$_{N(Local)}$ for S0s, 1.0$\pm$0.6 (0.1) is significantly smaller than the value of S$_{N(Local)}$ = 2.4$\pm$1.8
 (0.4) derived for ellipticals. Furthermore, there is weak evidence that S$_{N(local)}$ of early type S0s is slightly larger than that of the late types. Thus we conclude that specific frequency is a function of the Hubble type of the host galaxy.

For a more detailed description of our results we refer the reader to the 
recently published and forthcoming accepted paper Kundu \& Whitmore (2001a) and
 (2001b) respectively.

\end{document}